\documentclass[preprint,12pt]{elsarticle}

\usepackage{amssymb}

\def\th{\theta}

\def\la{\lambda}

\def\si{\sigma}

\def\ph{\phi}

\def\ch{\chi}

\def\om{\omega}

\def\De{\Delta}

\def\nue{\nu_e}
\def\numu{\nu_\mu}
\def\nubar{\bar\nu}
\def\nuebar{\bar\nu_e}
\def\numubar{\bar\nu_\mu}

\def\fr#1#2{\frac{#1}{#2}}

\def\lsim{\mathrel{\rlap{\lower4pt\hbox{\hskip1pt$\sim$}}
    \raise1pt\hbox{$<$}}}
\def\gsim{\mathrel{\rlap{\lower4pt\hbox{\hskip1pt$\sim$}}
    \raise1pt\hbox{$>$}}}

\newcommand{\beq}{\begin{eqnarray}}
\newcommand{\eeq}{\end{eqnarray}}
\def\to{\rightarrow}

\def\no{\nonumber}

\def\aL{(a_L)}
\def\cL{(c_L)}

\def\As#1{({\cal A}_s)_{#1}}
\def\Ac#1{({\cal A}_c)_{#1}}
\def\Bs#1{({\cal B}_s)_{#1}}
\def\Bc#1{({\cal B}_c)_{#1}}
\def\C#1{({\cal C})_{#1}}

\def\nh^#1{{\hat N}^{#1}}
\def\indxn{{e\mu}}

\def\mF_#1{{\cal F}_{#1}} 

\def\MBosc1POT{5.58\times 10^{20}}
\def\evki{146,070}

\def\evkib{47,466}

\def\tsid{86164.1}
\def\tgmt{86400.0}

\def\MBch{48.2}
\def\MBth{89.8}
\def\MBph{180.0}

\def\Dcsqnu{26.9}
\def\Dcsqnubar{3.0}

\def\lowEnu{0.36}\def\lowEnubar{0.34}
\def\oscEnu{0.82}\def\oscEnubar{0.78}
\def\totEnu{0.71}\def\totEnubar{0.60}

\def\AsAcnu{0.8}
\def\AsAcnb{1.1}

\def\np{6.46\times 10^{20}}
\def\nln{544}\def\nlb{409.8}\def\nlt{23.3}\def\nly{38.3}
\def\non{420}              
\def\ntn{964}
\def\ap{5.66\times 10^{20}}
\def\aln{119}
\def\aon{122}
\def\atn{241}\def\atb{200.7}\def\att{15.5}\def\aty{14.3}

\def\nlrt{0.6}\def\nlry{0.9}\def\nlrl{4.2}\def\nlrva{3.1}
\def\nlst{0.9}\def\nlsy{0.3}\def\nlsl{3.3}\def\nlsva{0.6}
\def\nlct{0.9}\def\nlcy{0.4}\def\nlcl{4.0}\def\nlcva{0.4}

\def\nlbk{0.13}\def\Dnlbk{13}

\def\nlgk{0.42}
\def\nobk{0.64}

\def\nogk{0.81}
\def\ntbk{0.14}

\def\ntgk{0.64}

\def\atrt{0.8}\def\atry{0.1}\def\atrl{2.6}\def\atrva{0.1}
\def\atst{1.3}\def\atsy{0.5}\def\atsl{3.9}\def\atsva{2.4}
\def\atct{1.2}\def\atcy{0.4}\def\atcl{3.7}\def\atcva{2.1}

\def\albk{0.15}

\def\algk{0.62}
\def\aobk{0.39}

\def\aogk{0.79}
\def\atbk{0.08}\def\Datbk{8}

\def\atgk{0.69}

\journal{Physics Letters B}

\begin{document}

\begin{frontmatter}



\title{Test of Lorentz and CPT violation with Short Baseline Neutrino 
Oscillation Excesses}


\author{
A.~A. Aguilar-Arevalo$^{14}$, C.~E.~Anderson$^{19}$,
	A.~O.~Bazarko$^{16}$, S.~J.~Brice$^{8}$, B.~C.~Brown$^{8}$,
        L.~Bugel$^{13}$, J.~Cao$^{15}$, 
        L.~Coney$^{6}$\footnote{Present address: University of California; Riverside, CA 92521},
        J.~M.~Conrad$^{13}$, D.~C.~Cox$^{10}$, A.~Curioni$^{19}$, R.~Dharmapalan$^{1}$,
        Z.~Djurcic$^{2}$, D.~A.~Finley$^{8}$, B.~T.~Fleming$^{19}$,
        R.~Ford$^{8}$, F.~G.~Garcia$^{8}$,
        G.~T.~Garvey$^{11}$, J.~Grange$^{9}$, C.~Green$^{8,11}$, J.~A.~Green$^{10,11}$,
        T.~L.~Hart$^{5}$, E.~Hawker$^{4,11}$, W.~Huelsnitz$^{11}$,
        R.~Imlay$^{12}$, R.~A. ~Johnson$^{4}$, G.~Karagiorgi$^{13}$, P.~Kasper$^{8}$, 
        T.~Katori$^{10,13}$\footnote{Corresponding author. Teppei Katori: katori@fnal.gov}, 
        T.~Kobilarcik$^{8}$, I.~Kourbanis$^{8}$, S.~Koutsoliotas$^{3}$, E.~M.~Laird$^{16}$,
        S.~K.~Linden$^{19}$,J.~M.~Link$^{18}$, Y.~Liu$^{15}$,
        Y.~Liu$^{1}$, W.~C.~Louis$^{11}$,
        K.~B.~M.~Mahn$^{6}$, W.~Marsh$^{8}$, C.~Mauger$^{11}$,
	V.~T.~McGary$^{13}$, G.~McGregor$^{11}$,
        W.~Metcalf$^{12}$, P.~D.~Meyers$^{16}$,
        F.~Mills$^{8}$, G.~B.~Mills$^{11}$,
        J.~Monroe$^{6}$, C.~D.~Moore$^{8}$, J.~Mousseau$^{9}$, R.~H.~Nelson$^{5}$,
	P.~Nienaber$^{17}$, J.~A.~Nowak$^{12}$,
	B.~Osmanov$^{9}$, S.~Ouedraogo$^{12}$, R.~B.~Patterson$^{16}$,
        Z.~Pavlovic$^{11}$, D.~Perevalov$^{1,8}$, C.~C.~Polly$^{8}$, E.~Prebys$^{8}$,
        J.~L.~Raaf$^{4}$, H.~Ray$^{9}$, B.~P.~Roe$^{15}$,
	A.~D.~Russell$^{8}$, V.~Sandberg$^{11}$, R.~Schirato$^{11}$,
        D.~Schmitz$^{8}$, M.~H.~Shaevitz$^{6}$, F.~C.~Shoemaker$^{16}$\footnote{deceased},
        D.~Smith$^{7}$, M.~Soderberg$^{19}$,
        M.~Sorel$^{6}$\footnote{Present address: IFIC, Universidad de Valencia and CSIC, Valencia 46071, Spain},
        P.~Spentzouris$^{8}$, J.~Spitz$^{19}$, I.~Stancu$^{1}$,
        R.~J.~Stefanski$^{8}$, M.~Sung$^{12}$, H.~A.~Tanaka$^{16}$,
        R.~Tayloe$^{10}$, M.~Tzanov$^{12}$,
        R.~G.~Van~de~Water$^{11}$, 
	M.~O.~Wascko$^{12}$\footnote{Present address: Imperial College; London SW7 2AZ, United Kingdom},
	 D.~H.~White$^{11}$,
        M.~J.~Wilking$^{5}$, H.~J.~Yang$^{15}$,
        G.~P.~Zeller$^{8}$, E.~D.~Zimmerman$^{5}$ \\
\smallskip
(The MiniBooNE Collaboration)
\smallskip
}
\smallskip
\smallskip
\address{
$^1$University of Alabama; Tuscaloosa, AL 35487 \\
$^2$Argonne National Laboratory; Argonne, IL 60439 \\
$^3$Bucknell University; Lewisburg, PA 17837 \\
$^4$University of Cincinnati; Cincinnati, OH 45221\\
$^5$University of Colorado; Boulder, CO 80309 \\
$^6$Columbia University; New York, NY 10027 \\
$^7$Embry Riddle Aeronautical University; Prescott, AZ 86301 \\
$^8$Fermi National Accelerator Laboratory; Batavia, IL 60510 \\
$^9$University of Florida; Gainesville, FL 32611 \\
$^{10}$Indiana University; Bloomington, IN 47405 \\
$^{11}$Los Alamos National Laboratory; Los Alamos, NM 87545 \\
$^{12}$Louisiana State University; Baton Rouge, LA 70803 \\
$^{13}$Massachusetts Institute of Technology; Cambridge, MA 02139 \\
$^{14}$Instituto de Ciencias Nucleares, Universidad Nacional Aut\'onoma de M\'exico, D.F. 04510, M\'exico \\
$^{15}$University of Michigan; Ann Arbor, MI 48109 \\
$^{16}$Princeton University; Princeton, NJ 08544 \\
$^{17}$Saint Mary's University of Minnesota; Winona, MN 55987 \\
$^{18}$Virginia Polytechnic Institute \& State University; Blacksburg, VA 24061 \\
$^{19}$Yale University; New Haven, CT 06520\\
 }

\begin{abstract}
The sidereal time dependence of
MiniBooNE $\nue$ and $\nuebar$ appearance data are analyzed 
to search for evidence of Lorentz and CPT violation. 
An unbinned Kolmogorov-Smirnov test shows 
both the $\nue$ and  $\nuebar$ appearance data are compatible with 
the null sidereal variation hypothesis to more than 5\%.  
Using an unbinned likelihood fit 
with a Lorentz-violating oscillation model 
derived from the Standard Model Extension (SME) 
to describe any excess events over background, 
we find that the $\nue$ appearance data prefer 
a sidereal time-independent solution, 
and the $\nuebar$ appearance data slightly prefer 
a sidereal time-dependent solution. 
Limits of order $10^{-20}$~GeV are 
placed on combinations of SME coefficients. 
These limits give the best limits on certain SME coefficients for 
$\numu\to\nue$ and $\numubar\to\nuebar$ oscillations. 
The fit values and limits of combinations of 
SME coefficients are provided.
\end{abstract}


\end{frontmatter}



\section{Introduction of Loremtz violation}

Violation of Lorentz invariance and CPT symmetry is a 
predicted phenomenon of Planck scale physics, 
especially with a spontaneous violation~\cite{SLSB}, 
and it does not require any modifications 
in quantum field theory or general relativity. 
Since neutrino oscillation experiments are natural interferometers, 
they can serve as sensitive probes of space-time structure. 
Neutrino oscillations have the potential to 
provide the first experimental evidence for Lorentz and CPT violation 
through evidence of oscillations that deviate from 
the standard $L/E$ dependence~\cite{Coleman}, 
or that show sidereal time dependent oscillations 
as a consequence of a preferred direction in the universe~\cite{KM1}.

In this paper, we test the MiniBooNE $\numu\to\nue$ and 
$\numubar\to\nuebar$ oscillation data~\cite{MB_nu,MB_antinu} 
for the presence of a Lorentz violation signal. 
Similar analyses have been performed in other oscillation experiments, 
including LSND~\cite{LSND_LV}, MINOS~\cite{MINOS_LV}, and IceCube~\cite{IceCube_LV}. 
Naively, experiments with longer baselines and higher energy neutrinos 
would be expected to have better sensitivity to Lorentz violation, 
because small Lorentz violating terms are more prominent at high energy, 
where neutrino mass terms are negligible. 
However, some Lorentz violating neutrino oscillation models mimic the standard 
massive neutrino oscillation energy dependence~\cite{KM2}. 
Then, in this case, the signal may only be seen in 
sidereal variations 
of oscillation experiments.

\section{MiniBooNE experiment}

MiniBooNE is a $\nue$ ($\nuebar$) appearance short baseline neutrino oscillation experiment at Fermilab. 
Neutrinos are created by the Booster Neutrino Beamline (BNB), 
which produces a 93\% (83\%) pure $\numu$ ($\numubar$) beam in neutrino (anti-neutrino) mode, 
determined by the polarity of the magnetic focusing horn. 
The MiniBooNE Cherenkov detector, 
a 12.2~m diameter sphere filled with mineral oil, 
is used to detect charged particles from neutrino interactions and 
is located 541~m from the neutrino production target. 
It is equipped with 1,280 8'' PMTs in an optically separated inner volume, 
and 240 8'' veto PMTs in an outer veto region. 
Details  of the detector and the BNB can be found elsewhere~\cite{MB_beam,MB_detec}. 
Charged leptons created by neutrino interactions in the detector produce Cherenkov photons, 
which are used to reconstruct charged particle tracks~\cite{MB_recon}. 
The measured angle and kinetic energy of charged leptons from neutrino interactions 
are used to reconstruct the neutrino energy, $E_\nu^{QE}$, for each event, 
under the assumption that the target nucleon is at rest inside the nucleus and 
the interaction type is a charged current quasielastic (CCQE) interaction~\cite{MB_CCQEPRL}.  

For this analysis, we use the background and error estimates 
from~\cite{MB_nu_public} (neutrino mode) 
and~\cite{MB_antinu_public} (antineutrino mode).
For neutrino mode, 
data from $\np$  protons on target (POT) are used. 
An excess in the ``low energy" region (200$<E_\nu^{QE}$(MeV)$<$475) was observed, 
with $\nln$ events reported as compared to the prediction, 
$\nlb\pm\nlt$(stat.)$\pm\nly$(syst.). 
Interestingly, this excess does not show the expected $L/E$ 
energy dependence of a simple two massive neutrino oscillation model. 
Additionally, it is not consistent with 
the energy region expected for the ``LSND'' signal~\cite{LSND_osc}. 
For the anti-neutrino mode analysis ($\ap$POT), 
MiniBooNE observed a small excess in the low energy region, 
and an excess in the region 475$<E_\nu^{QE}$(MeV)$<$1300. 
The excess in this ``high energy'' region is found 
to be consistent with the LSND signal, 
assuming a two massive neutrino hypothesis, 
but remains statistically marginal.  
In the ``combined" region (200$< E_\nu^{QE}$(MeV)$<$1300), 
MiniBooNE observed $\atn$ $\nuebar$ candidate events as compared to the prediction, 
$\atb\pm\att$(stat.)$\pm\aty$(syst.). 

Although the conflict between MiniBooNE neutrino and anti-neutrino mode results 
can be resolved in models without CPT violation~\cite{Kopp}, 
CPT violation is a viable option. 
Since CPT violation necessarily implies violation of Lorentz invariance 
within interactive quantum field theory~\cite{Greenberg}, 
we are in a well-motivated position to search 
for Lorentz and CPT violation using the MiniBooNE data. 
In fact,  proposed models motivated by Lorentz violation~\cite{tandem,puma} 
can already accommodate world data including the MiniBooNE and LSND excesses  
with a small number of free parameters. 
Evidence for sidereal variation in the MiniBooNE excesses would provide 
a distinctive direct signal of Lorentz violation.

\section{Analysis}

We use the Standard Model Extension (SME) formalism for 
the general search for Lorentz violation~\cite{SME}. 
The SME is an effective quantum field theory, 
and the minimum extension of the Standard Model 
including particle Lorentz and CPT violation~\cite{SME}. 
A variety of data have been analyzed under this formalism~\cite{CPT10}, 
including neutrino oscillations~\cite{LSND_LV,MINOS_LV,IceCube_LV}. 
In the SME formalism for neutrinos, 
the evolution of a neutrino can be described by 
an effective Hamiltonian~\cite{KM1},
\beq
(h^\nu_{\rm{eff}})_{ab}\sim
\fr{1}{E}[(a_L)^{\mu} p_{\mu}-(c_L)^{\mu\nu}p_{\mu}p_{\nu}]_{ab}~.
\label{eq:hamiltonian} 
\eeq
Here, $E$ and $p_{\mu}$ are the energy and the 4-momentum of a neutrino, 
and $\aL^{\mu}_{ab}$ and $\cL^{\mu\nu}_{ab}$ 
are CPT-odd and CPT-even SME coefficients in the flavor basis. 
Under the assumption that the baseline is short compared to the oscillation length~\cite{KM3}, 
the $\numu\to\nue$ oscillation probability takes the form, 
\small
\beq
P\simeq\frac{L^2}{(\hbar c)^2}|\C{\indxn}+\As{\indxn}\sin\om_\oplus T_\oplus
+\Ac{\indxn} \cos\om_\oplus  T_\oplus\no\\
+\Bs{\indxn} \sin2\om_\oplus T_\oplus 
+\Bc{\indxn} \cos2\om_\oplus T_\oplus|^2.\label{eq:SBA}
\eeq
\normalsize
This probability is a function of sidereal time $T_\oplus$. 
Four parameters, 
$\As{\indxn}$, 
$\Ac{\indxn}$, 
$\Bs{\indxn}$, and 
$\Bc{\indxn}$ are sidereal time dependent, 
and $\C{\indxn}$ is a sidereal time independent parameter. 
We use a baseline distance of $L=$522.6~m, 
where the average pion decay length is subtracted from 
the distance between the neutrino production target and detector.  
And $\om_\oplus$ is the sidereal time angular frequency described shortly. 
These parameters are expressed in terms of SME coefficients 
and directional factors~\cite{KM3}. 
The same formula describes the $\numubar\to\nuebar$ oscillation probability by 
switching the signs of the CPT-odd SME coefficients. 
We neglect the standard neutrino mass term, $m^2_{\indxn}/E\ll10^{-20}$~GeV, 
which is well below our sensitivity which is discussed later.


For this analysis, 
we convert the standard GPS time stamp for each event to local solar time (period $\tgmt$~sec) 
and sidereal time (period $\tsid$~sec).  
We then define the local solar time angular frequency $\om_\odot=\frac{2\pi}{\tgmt}$~(rad/s) and 
the sidereal time angular frequency $\om_\oplus=\frac{2\pi}{\tsid}$~(rad/s). 
The time origin can be arbitrary, 
but we follow the standard convention with a Sun-centered coordinate system~\cite{LSND_LV}.
We choose a time-zero of 58~min after an autumnal equinox of 2002 (September 23, 04:55GMT), 
so that this serves not only as the sidereal time-zero, 
but also as the solar time-zero since Fermilab is on the midnight point at this time. 
The local coordinates of the BNB are specified by three angles~\cite{KM3},  
colatitude $\ch=\MBch^\circ$, 
polar angle $\th=\MBth^\circ$, 
and azimuthal angle $\ph=\MBph^\circ$.

\begin{figure}[t!]
\includegraphics[width=\columnwidth]{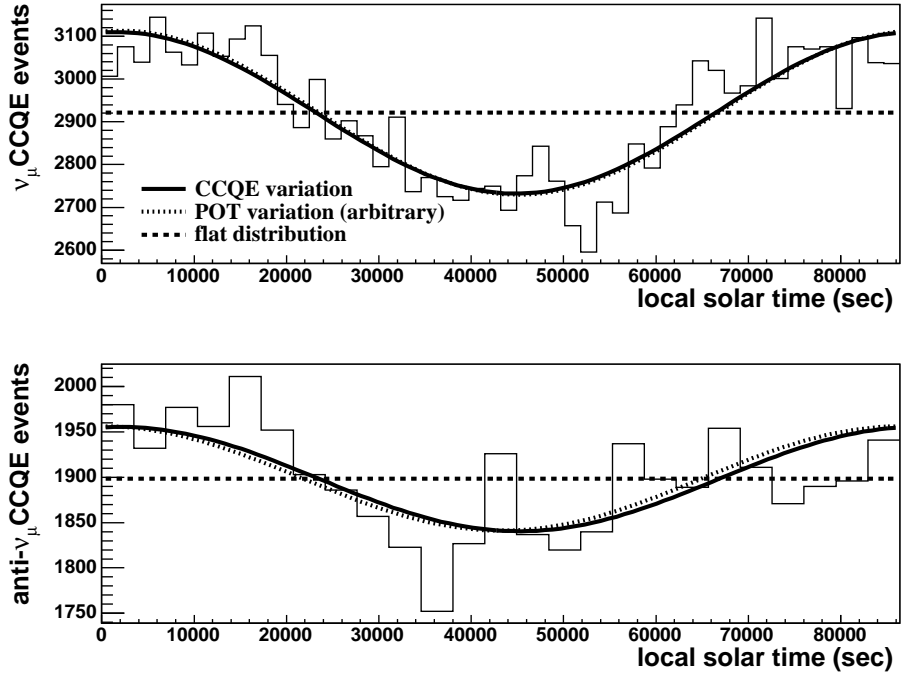}
\vspace{-2mm}
\caption{
The top (bottom) plot shows the $\numu$CCQE  ($\numubar$CCQE) local solar time distribution. 
The solid curves are fit functions 
extracted from the CCQE event distributions, 
and the dotted curves are from the POT distributions (arbitrary units) 
during the same period of data taking. 
The dashed line shows a flat distribution.
}
\label{fig:tdist}
\end{figure}

Any time dependent background variation, 
such as the time variation of detector and BNB systematics, are important. 
To evaluate these, 
we use  our high statistics CCQE samples (Figure~\ref{fig:tdist}). 
These data are from our 
$\numu$CCQE double differential cross section measurement sample~\cite{MB_CCQEPRD} 
composed of $\evki$ events ($\MBosc1POT$ POT), 
and our $\numubar$CCQE candidate sample~\cite{MB_ANTICCQE} 
composed of $\evkib$ events ($\ap$ POT). 
The $\numu$($\numubar$)CCQE local solar time distribution exhibits $\pm$6(3)\% variation. 

The same variation in local solar time is observed in the POT for both 
neutrino and anti-neutrino mode data taking periods. 
Therefore, the POT variation is the dominant time dependent systematic error. 
The amplitude is negligible in $\numubar$CCQE sidereal time distribution, 
however, it persists in $\sim$3\% variations in $\numu$CCQE sidereal time distribution. 
We evaluate the impact of this variation on our analysis by correcting 
POT variation event by event in $\nue$ ($\nuebar$) candidate data. 
It turned out the correction only has a negligible effect. 
Thus we decided to use unweighted events.  
This also simplifies the unbinned likelihood function used in later analysis.  
Figures~\ref{fig:nu_3F_lowE} and~\ref{fig:nub_3F_totE} 
show the $\nue$ and $\nuebar$ oscillation candidates 
sidereal time distributions both with and without the POT correction.
These plots verify that time-dependent systematics are negligible in this analysis.

To check for a general deviation from a flat distribution 
(null sidereal variation hypothesis),  
we perform an unbinned Kolmogorov-Smirnov test (K-S test)~\cite{NR} 
as a statistical null hypothesis test 
for both the $\nue$ and $\nuebar$ samples. 
The K-S test is suitable in our case because it is sensitive to runs in distributions, 
which may be a characteristic feature of the sidereal time dependent hypothesis. 
Table~\ref{tab:stattest} gives the result. 
The K-S test is applied to the low energy, high energy, and combined regions, 
for both neutrino, and anti-neutrino mode data. 
To investigate the time dependent systematics, 
we also apply the K-S test to the local solar time distribution. 
The test shows none of the twelve samples has less than 5\% compatibility ($\sim 2\si$), 
which we chose as a benchmark prior to the analysis. 
Hence, all samples are compatible with the null sidereal variation hypothesis. 
Interestingly, the sidereal time distributions tend to show lower compatibility with a 
flat hypothesis, 
but not by a statistically significant amount. 
These results indicate that any sidereal variation extracted from our data, 
discussed below, 
is not expected to be statistically significant. 

\begin{table}
\begin{center}
\small
\begin{tabular}{lcccccc}
\hline
\hline
 & \multicolumn{2}{c}{~~~low energy~~~} 
 & \multicolumn{2}{c}{~~~~high energy~~~} 
 & \multicolumn{2}{c}{~~~~combined~~~~}\\
&solar~&~sidereal~&~~~solar~&~sidereal~&~~~solar~&~sidereal\\
\hline
\multicolumn{7}{c}{Neutrino mode} \\
$<E_\nu>$
& \multicolumn{2}{c}{$\lowEnu$~GeV} 
& \multicolumn{2}{c}{$\oscEnu$~GeV} 
& \multicolumn{2}{c}{$\totEnu$~GeV}\\
\#evt 
& \multicolumn{2}{c}{\nln} 
& \multicolumn{2}{c}{\non} 
& \multicolumn{2}{c}{\ntn}\\
$P(\mathrm{KS})$&\nlgk&\nlbk&\nogk&\nobk&\ntgk&\ntbk\\
\hline
\multicolumn{7}{c}{Anti-neutrino mode} \\
$<E_\nu>$
& \multicolumn{2}{c}{$\lowEnubar$~GeV} 
& \multicolumn{2}{c}{$\oscEnubar$~GeV} 
& \multicolumn{2}{c}{$\totEnubar$~GeV}\\
\#evt 
& \multicolumn{2}{c}{\aln} 
& \multicolumn{2}{c}{\aon}
& \multicolumn{2}{c}{\atn}\\
$P(\mathrm{KS})$&\algk&\albk&\aogk&\aobk&\atgk&\atbk\\
\hline
\hline
\end{tabular}
\end{center}
\caption{\label{tab:stattest}
A summary of K-S test results on the
sidereal and local solar time distributions. 
The top table is for $\nue$ candidate data, 
and the bottom table is for  $\nuebar$ candidate data. 
The three rows show the average neutrino energy of each sample, 
number of events, and the K-S test compatibility with the null hypothesis. 
The test is performed in three energy regions, 
and for both solar local time and sidereal time distributions.
}
\end{table}


To fit the data with the sidereal time-dependent model, 
we use a generalized unbinned maximum likelihood method~\cite{Lyons}. 
This method finds the best fit model parameters 
by fitting data with a log likelihood function $\ell$. 
It is suitable for our analysis because this method has 
the highest statistical power for a low statistics sample. In this method, 
the log likelihood function $\ell$ is constructed by adding $\ell_i$ from each event.  
After dropping all constants, $\ell_i$ has the following expression,  
\small
\beq
\ell_i=-\fr{1}{N}(\mu_s+\mu_b)+ln[\mu_s \mF_s^i + \mu_b \mF_b^i]
-\frac{1}{2N}\left(\fr{\mu_b-\overline{\mu_b}}{\si_b}\right)^2.
\label{eq:ell} 
\eeq
\normalsize
Here, 
$N$ is the number of observed candidate events, 
$\mu_s$ is the predicted number of signal events 
which is given by the time integral of Eq.~\ref{eq:SBA} 
together with the estimated efficiency, 
$\mu_b$ is the predicted number of background events,  
$\mF_s$ is the probability density function (PDF) for the signal 
and is a function of sidereal time and the fitting parameters 
(Eq.~\ref{eq:SBA} with proper normalization), 
$\mF_b$ is the PDF for the background, 
$\si_b$ is the $1\si$ error on the predicted background, and 
$\overline{\mu_b}$ is the central value of the predicted total background events. 
Two sources contribute equally to the background:  
intrinsic beam background and mis-identification (mainly $\pi^{\circ}$s). 
Their total variation is assigned as the systematic error, 
assumed to be time independent. 
Details can be found in~\cite{MB_nu,MB_antinu}.  
The parameter space is scanned (grid search method) to find the largest $\ell$, 
or the maximum log likelihood (MLL) point,  
and this MLL point provides the combination of the best fit (BF) parameters. 
The log likelihood function includes six parameters, 
five that are functions of SME coefficients, and one for the background. 
However the background term is constrained within a $\pm1\si$ range. 
Neither the neutrino nor the anti-neutrino mode data allow us to extract errors 
if we fit all five parameters at once, due to the  high correlation of parameters. 
Therefore, we set $\Bs\indxn$ and $\Bc\indxn$ to zero
and concentrate on three parameters ($\C\indxn$, $\As\indxn$, and $\Ac\indxn$) 
for the uncertainty estimates. 
Since the five parameter fit is quantitatively similar to the three parameter fit, 
we will focus the discussion on the results on the three parameter fits. 
This three parameter fit also corresponds to the case 
with only CPT-odd SME coefficients~\cite{KM3}.

\section{Results}

\begin{figure}[t!]
\includegraphics[width=\columnwidth]{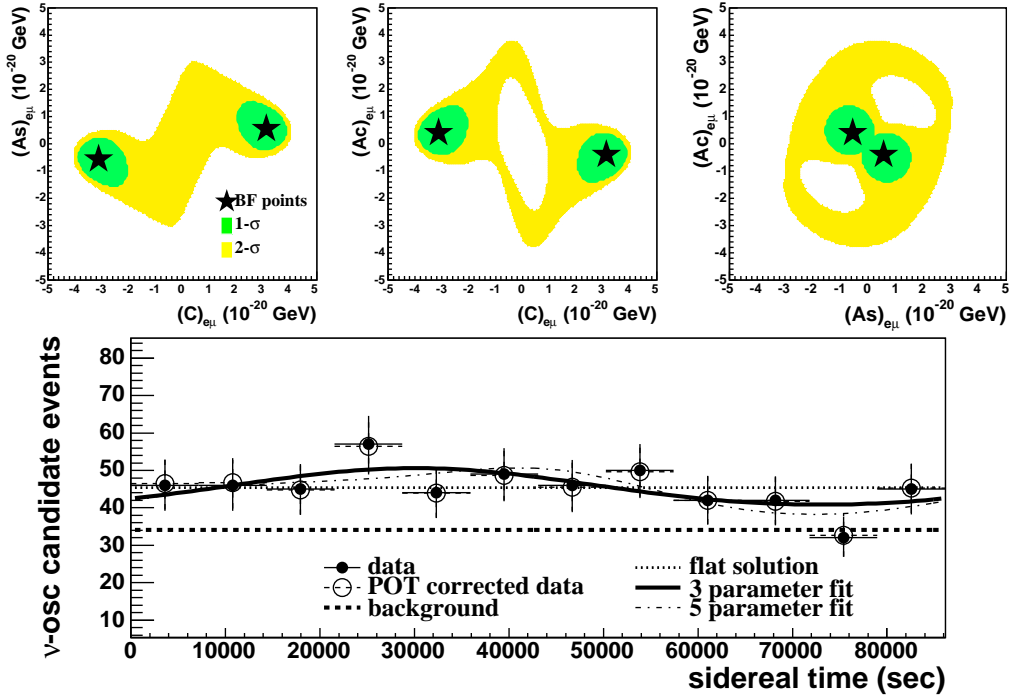}
\vspace{-2mm}
\caption{(color online) 
Three parameter fit results for the neutrino mode low energy region.
The top three plots show the projection of 
three dimensional parameter space.  
The dark (light) shaded area shows the $1\si$ ($2\si$) contour in each projection. 
The stars show the BF points. 
The bottom plot shows the curves corresponding to the 
flat solution (dotted),  
three parameter fit (solid),   
and five parameter fit (dash-dotted), together with binned data (solid marker). 
The POT corrected data are also shown in open circle marker. 
Here, the fitted background is shown as a dashed line, and the BF value is 1.00 
({\it i.e.}, equivalent to the central value of the predicted background).
}
\label{fig:nu_3F_lowE}
\end{figure}

Figure~\ref{fig:nu_3F_lowE} shows the neutrino mode low energy region fit results. 
The top three plots show the three projections of three dimensional parameter space. 
Because of the square of fitting parameters in the PDF,
the BF point has a sign ambiguity and is always duplicated. 
The  $1\si$ and $2\si$ contours are formed from a 
constant slice of the log likelihood function in the three dimensional parameter space. 
To avoid under coverage, these slices are expanded until they enclose 68\% ($1\si$) or 95\% ($2\si$) of 
BF points for the three parameter fit of simulated, or ``fake'', data sets
with the signal. 
Note that because fitting parameters are not linear in the PDF, 
twice the $1\si$ error does not yield the $2\si$ error. 


A null sidereal variation hypothesis, or a flat solution,  
is equivalent to a three or five parameter fit solution where 
only the $\C\indxn$ parameter is nonzero. 
The fit to neutrino data favors
a nonzero solution only for the $\C\indxn$ term.
The bottom plot in Figure~\ref{fig:nu_3F_lowE} 
shows data plotted against curves corresponding to the flat solution and the best
fits for three and five parameter functions. 
Since all three curves are close to each other, 
the solution of neutrino mode is dominated by the sidereal time-independent component. 
To find the significance of time dependence over the flat distribution, 
fake data sets without a signal are formed where 
the $\nue$ candidate events are simulated without any time structure. 
The MLL difference between the three parameter fit 
and the flat solution is used to form a $\De\ch^2$, 
and the expected $\De\ch^2$ distribution is determined by testing 500 random
distributions from the fake data sets.  
This test shows that there is a $\Dcsqnu$\% chance that a
random distribution of $\nue$ candidate events 
would yield a $\De\ch^2$ value equal to, 
or greater than,
the value observed for the data. 
This result is consistent with the sensitivity of this experiment. 
We estimate our sensitivity to the limitted case.  
First, a $2\si$ threshold is set from this $\De\ch^2$ distribution. 
Then, time dependent amplitudes were incrementally increased 
in the model until the $2\si$ threshold was exceeded. 
When we assume $\C\indxn\neq 0$ and $\As\indxn=\Ac\indxn\neq 0$ 
the $2\si$ discovery threshold of sidereal time dependent amplitudes from 
$\nue$($\nuebar$) candidate data statistics are 
$\AsAcnu$($\AsAcnb$)$\times 10^{-20}$~GeV. 

\begin{table*}
\begin{center}
\footnotesize
\begin{tabular}{lcccc}
\hline
\hline
&$\nu-$mode BF&2$\si$ limit&$\nubar -$mode BF&2$\si$ limit\\
\hline
$|\C\indxn|$ &~$\nlrva\pm\nlrt\pm\nlry$~&~$<\nlrl$~&~$\atrva\pm\atrt\pm\atry$~&~$<\atrl$\\
$|\As\indxn|$&~$\nlsva\pm\nlst\pm\nlsy$~&~$<\nlsl$~&~$\atsva\pm\atst\pm\atsy$~&~$<\atsl$\\
$|\Ac\indxn|$&~$\nlcva\pm\nlct\pm\nlcy$~&~$<\nlcl$~&~$\atcva\pm\atct\pm\atcy$~&~$<\atcl$\\
\hline
&\multicolumn{4}{c}{SME coefficients combination (unit $10^{-20}$~GeV)}\\
\hline
$|\C\indxn|$ &\multicolumn{4}{c}{
~~$\pm[\aL^T_\indxn+0.75\aL^Z_\indxn]-<E>[1.22\cL^{TT}_\indxn+1.50\cL^{TZ}_\indxn+0.34\cL^{ZZ}_\indxn]$}\\
$|\As\indxn|$&\multicolumn{4}{c}{
~~~~$\pm[0.66\aL^Y_\indxn]-<E>[1.33\cL^{TY}_\indxn+0.99\cL^{YZ}_\indxn]$}\\
$|\Ac\indxn|$&\multicolumn{4}{c}{
~~~~$\pm[0.66\aL^X_\indxn]-<E>[1.33\cL^{TX}_\indxn+0.99\cL^{XZ}_\indxn]$}\\
\hline
\hline
\end{tabular}
\end{center}
\caption{
The fit parameters for the neutrino mode low energy region and the anti-neutrino mode combined region. 
The BF points are the MLL points of the log likelihood function, 
here top rows from left to right, BF values, $1\si$ statistical, and systematic errors. 
The $2\si$ limits are also shown. 
Bottom rows show detailed expressions of each sidereal fit parameter in terms of SME parameters, 
and directional factors~\cite{KM3}. 
The upper (lower) sign of $\aL^\la_\indxn$ terms is applied for neutrino (anti-neutrino) results, 
due to the CPT-odd nature. 
The average neutrino energy ``$<E>$'' is $\lowEnu$~GeV for the neutrino mode low energy region 
and $\totEnubar$~GeV for the anti-neutrino mode combined region (Tab.~\ref{tab:stattest}).
}
\label{tab:largeSME}
\end{table*}



\begin{figure}[t!]
\includegraphics[width=\columnwidth]{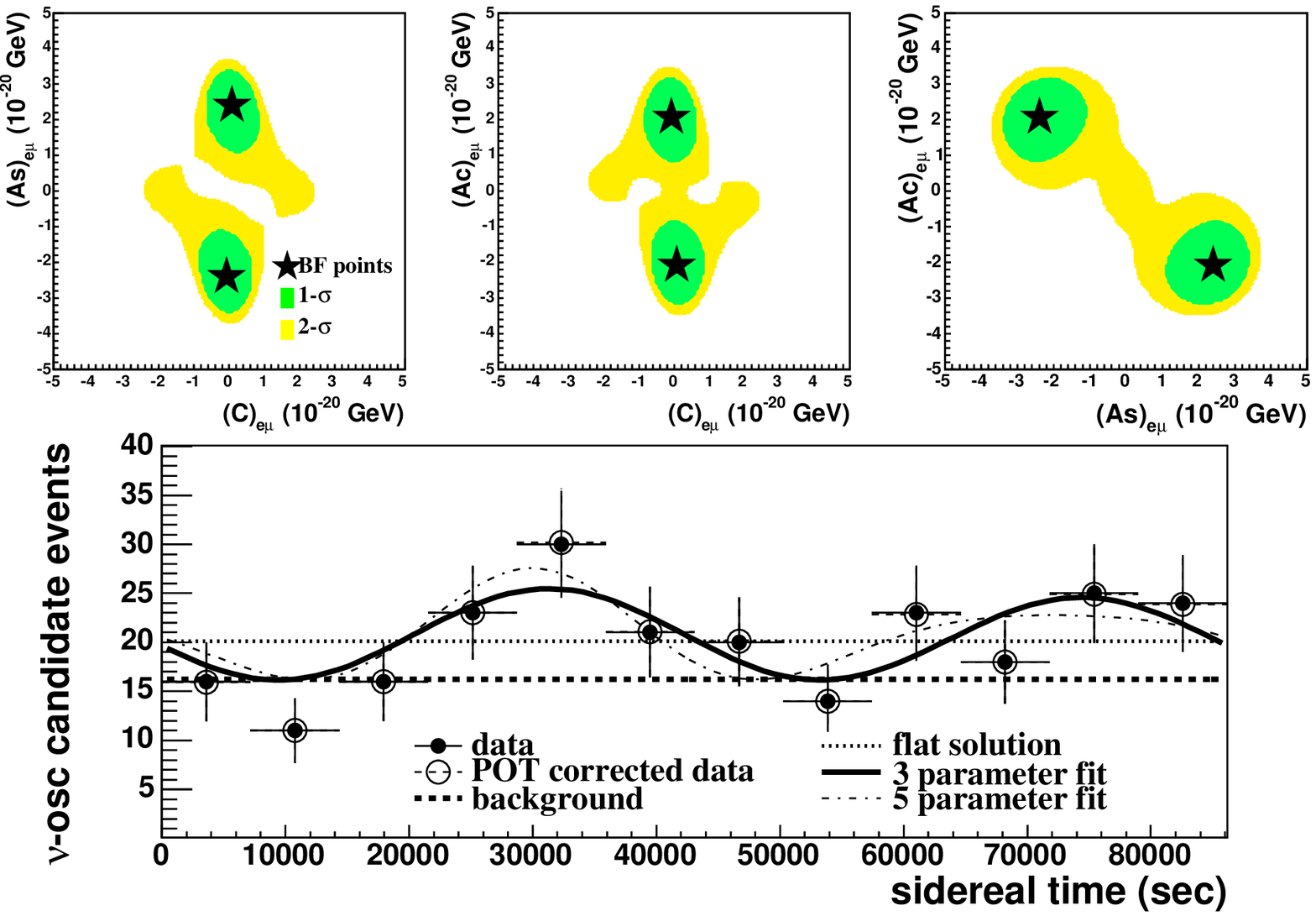}
\vspace{-2mm}
\caption{
Three parameter fit results for the anti-neutrino mode combined region. 
Notations are the same as Fig.~\ref{fig:nu_3F_lowE}. 
Here, the BF value for the fitted background is 0.97 
(3\% lower than the central value of the predicted background).
}
\label{fig:nub_3F_totE}
\end{figure}


Figure~\ref{fig:nub_3F_totE} shows the analogous fit results for 
the anti-neutrino mode combined energy region.  
Due to lower statistics, 
the combined region is used rather than dividing the data into two subsets. 
Unlike the neutrino mode low-energy region, 
the $\C\indxn$ parameter no longer significantly deviates from zero. 
The fit to anti-neutrino data favors
a non-zero solution for the $\As\indxn$ and $\Ac\indxn$ parameters at the nearly 2$\si$ level. 
Performing the same $\De\ch^2$ test as is outlined above results in 
only $\Dcsqnubar$\% of the random distributions from the $\nuebar$ candidate events 
having a $\De\ch^2$ value exceeding the value observed for the data. 
Note that this is consistent with the $\Datbk$\% compatibility with a flat hypothesis 
found with the K-S test (Tab.~\ref{tab:stattest}). 
 

Table~\ref{tab:largeSME} shows fit parameters for the neutrino mode low energy region 
and anti-neutrino mode combined region. 
All errors are estimated from $1\si$ contours of parameter space projections. 
Errors are asymmetric, but we choose the larger excursions as the symmetric errors for simplicity. 
The $2\si$ contours provide the limits. 
In principle, these fit parameters are complex numbers. 
Here, all parameters are assumed to be real. 
A naive estimation from Tab.~\ref{tab:largeSME} indicates  
possible SME coefficients to satisfy the MiniBooNE data are of order $10^{-20}$~GeV (CPT-odd), 
and $10^{-20}$ to $10^{-19}$ (CPT-even). 
However, these SME coefficients are too small to produce a visible effect for LSND~\cite{LSND_LV}. 
On the other hand, any SME coefficients extracted from LSND~\cite{LSND_LV} predict  
too large of a signal for MiniBooNE. 
Therefore, a simple picture using Lorentz violation 
to explain both data sets leaves some tension, 
and a mechanism to cancel the Lorentz violating effect 
at high energy~\cite{KM1,tandem,puma} is needed.

\section{summary}

In summary, we performed a sidereal time variation analysis 
for MiniBooNE $\nue$ and $\nuebar$ appearance candidate data. 
For the neutrino mode low energy region, 
K-S test statistics indicate the null hypothesis is compatible at the
$\Dnlbk$\% level, and the relative improvement in the likelihood between
the null hypothesis and the three parameter fit occurs 
$\Dcsqnu$\% of the time in random distributions from a null hypothesis. 
Analysis of the combined energy region in anti-neutrino mode results in a
K-S test that indicates a $\Datbk$\% compatibility with the null hypothesis, 
however the relative improvement in the likelihood between
the null hypothesis and the three parameter fit only occurs $\Dcsqnubar$\%
of the time in random distributions from a null hypothesis. 
The limits of fit parameters, $10^{-20}$~GeV, 
are consistent with Planck scale suppressed physics. 
This is the first sidereal variation test for an anti-neutrino beam of 
$\sim$1~GeV energy and $\sim$500~m base line. 
These limits are currently the best limits 
on the sidereal-time independent $\aL_\indxn$ and $\cL_\indxn$ SME coefficients. 
These limits can be significantly improved by long baseline $\nue$ ($\nuebar$) appearance 
experiments, such as T2K~\cite{T2K} and NOvA~\cite{NOvA}. 

\section*{acknowledgment}

This work was conducted with support from Fermilab, the U.S. Department of Energy,
the National Science Foundation and the Indiana University Center for Spacetime Symmetries.





\bibliographystyle{elsarticle-num}
\bibliography{<your-bib-database>}

\begin{thebibliography}{00}
\bibitem{SLSB}
V.~A.~Kosteleck\'y and S.~Samuel,
Phys. Rev. D 39 (1989) 683. 

\bibitem{Coleman}
S.~Coleman and S.~L.~Glashow,
Phys. Rev. D 59 (1999) 116008.

\bibitem{KM1}
V.~A.~Kosteleck\'y and M.~Mewes, 
Phys.~Rev.~D 69 (2004) 016005.

\bibitem{MB_nu}
A.~A.~Aguilar-Arevalo {\it et~al.}, 
Phys.~Rev.~Lett. 98 (2007) 231801; 
Phys.~Rev.~Lett. 102 (2009) 101802.

\bibitem{MB_antinu}
A.~A.~Aguilar-Arevalo {\it et~al.}, 
Phys.~Rev.~Lett. 103 (2009) 111801; 
Phys.~Rev.~Lett. 105 (2010) 181801.

\bibitem{LSND_LV}
L.~B.~Auerbach {\it et~al.}, 
Phys.~Rev.~D 72 (2005) 076004.

\bibitem{MINOS_LV}
P.~Adamson {\it et~al.}, 
Phys.~Rev.~Lett. 101 (2008) 151601;  
Phys.~Rev.~Lett. 105 (2010) 151601.

\bibitem{IceCube_LV}
R.~Abbasi {\it et~al.}, 
Phys.~Rev.~D 82 (2010) 112003.

\bibitem{KM2}
V.~A.~Kosteleck\'y and M.~Mewes, 
Phys.~Rev.~D 70 (2004) 031902 (R).

\bibitem{MB_beam}
A.~A.~Aguilar-Arevalo {\it et~al.}, 
Phys.~Rev.~D 79 (2009) 072002.

\bibitem{MB_detec}
A.~A.~Aguilar-Arevalo {\it et~al.}, 
Nucl.~Instrum.~Meth.~A 599 (2009) 28.

\bibitem{MB_recon} 
R.~B.~Patterson {\it et~al.}, 
Nucl.~Instrum.~Meth.~A 608 (2009) 206.

\bibitem{MB_CCQEPRL} 
A.~A.~ Aguilar-Arevalo {\it et~al.}, 
Phys.~Rev.~Lett. 100 (2008) 032301.

\bibitem{MB_nu_public}
\texttt{\normalsize http://www-boone.fnal.gov/for\_physicists/data\_release/lowe}

\bibitem{MB_antinu_public}
\texttt{\normalsize http://www-boone.fnal.gov/for\_physicists/data\_release/nuebar2010}

\bibitem{LSND_osc}
A.~Aguilar {\it et al.},
Phys.~Rev.~D 64 (2001) 112007.

\bibitem{Kopp}
J.~Kopp {\it et al}, 
Phys.~Rev.~Lett. 107 (2011) 091801.

\bibitem{Greenberg} 
O.~W.~Greenberg, 
Phys.~Rev.~Lett. 89 (2002) 231602.

\bibitem{tandem}
T.~Katori, V.~A.~Kosteleck\'y, and R.~Tayloe, 
Phys.~Rev.~D 74 (2006) 105009.

\bibitem{puma}
J.~S.~D\'iaz and V.~A.~Kosteleck\'y, 
Phys.~Lett.~B 700 (2011) 25; 
Phys.\ Rev.\ D 85 (2012) 016013.

\bibitem{SME}
D.~Colladay and V.~A.~Kosteleck\'y, 
Phys.~Rev.~D 55 (1997) 6760; 
Phys.~Rev.~D 58 (1998) 116002; 
V.~A.~Kosteleck\'y, 
Phys.~Rev.~D 69 (2004) 105009.

\bibitem{CPT10}
For recent reviews see,
for example, 
{\it CPT and Lorentz Symmetry V}, 
edited by V.~A.~Kosteleck\'y, 
(World Scientific, Singapore, 2011); 
V.~A.~Kosteleck\'y and N.~Russell, 
Rev.~Mod.~Phys.~83 (2011) 11.

\bibitem{KM3} 
V.~A.~Kosteleck\'y and M.~Mewes, 
Phys.~Rev.~D 70 (2004) 076002.

\bibitem{MB_CCQEPRD}
A.~A.~ Aguilar-Arevalo {\it et~al.},
Phys.~Rev.~D 81 (2010) 092005.

\bibitem{MB_ANTICCQE}
A.~A.~ Aguilar-Arevalo {\it et~al.}, 
Phys.~Rev.~D 84 (2011) 072005.

\bibitem{NR}
W.~H.~Press {\it et~al.}, 
{\it Numerical Recipes in C++: The Art of Scientific Computing}, 
(Cambridge University Press, New York, 2002).

\bibitem{Lyons} 
L.~Lyons, 
{\it Statistics for Nuclear and Particle Physicists}, 
(Cambridge University Press, New York, 1989).

\bibitem{T2K}
K.~Abe {\it et al.}, 
Phys. Rev. Lett. 107 (2011) 041801.

\bibitem{NOvA}
D.~S.~Ayres {\it et al.}, 
arXiv:hep-ex/0503053.

\end{thebibliography}



\end{document}